


\headline={\ifnum\pageno=1\firstheadline\else
\ifodd\pageno\rightheadline \else\leftheadline\fi\fi}
\def\firstheadline{\hfil}
\def\rightheadline{\hfil}
\def\leftheadline{\hfil}
	\footline={\ifnum\pageno=1\firstfootline\else\otherfootline\fi}
\def\firstfootline{\rm\hss\folio\hss}
\def\otherfootline{\hfil}

\font\twelvebf=cmbx10 scaled\magstep 1
\font\twelverm=cmr10 scaled\magstep 1
\font\twelveit=cmti10 scaled\magstep 1

\font\tenbf=cmbx10
\font\tenrm=cmr10
\font\tenit=cmti10

\parindent=1.5pc
\hsize=6.0truein
\vsize=8.5truein
\nopagenumbers

\centerline{\tenbf HARD THERMAL LOOPS, CHERN-SIMONS THEORY AND}
\baselineskip=22pt
\centerline{\bf THE QUARK-GLUON PLASMA}
\baselineskip=16pt
\vglue 0.8cm
\centerline{\tenrm V.P.NAIR }
\baselineskip=13pt
\centerline{\tenit Physics Department, City College of the CUNY }
\baselineskip=12pt
\centerline{\tenit New York, New York 10031, U.S.A.}
\centerline{\tenit and}
\centerline{\tenit Physics Department, Columbia University}
\centerline{\tenit New York, New York 10027, U.S.A.}
\vglue 0.8cm
\centerline{\tenrm ABSTRACT}
\vglue 0.3cm
{\rightskip=3pc
 \leftskip=3pc
 \tenrm\baselineskip=12pt\noindent
The generating functional for hard thermal loops in Quantum
Chromodynamics is important in setting up a resummed perturbation
theory.
I review how this functional is related to the eikonal for a
Chern-Simons gauge
theory, and using an auxiliary field, to the gauged
Wess-Zumino-Novikov-Witten action.
The induced current due to hard thermal loops, properly incorporating
damping effects, is also briefly discussed.
\vglue 0.6cm}
\footnote{ Invited talk at the Third Workshop on Thermal Field Theories,
Banff, Canada, August 1993.}
\vfil
\twelverm
\baselineskip=14pt
\vglue 0.3cm
In this talk, I shall be discussing how Chern-Simons theory can help us
understand and provide an elegant mathematical framework for the hard
thermal loops of Quantum Chromodynamics (QCD) and some of the physical
phenomena associated with the quark-gluon plasma. Some of the work I
describe was done in collaboration with R.Efraty $^1$ and some in
collaboration with Professor R.Jackiw $^2$.

This is an audience that is very familiar with the concept of hard
thermal loops; nevertheless, let me begin by briefly recalling what they
are. In QCD, hard thermal loops are thermal one-loop Feynman digrams for
which the momenta on the external lines are of the order of $gT$ or
less, where $g$ is the coupling constant and $T$ is the temperature and
for which the loop-momentum is of the order of $T$ or higher. As is well
known, the evaluation of hard thermal loops is necessary for the
Braaten-Pisarski resummation procedure $^{3,4}$.
Hard thermal loops are also part
of the effective action for long wavelength excitations of the plasma.
Let me denote by $\Gamma [A]$ the generating functional for hard thermal
loops with external gluon lines, $A_\mu$ being the gauge potential of
the external lines. ( Hard thermal loops with external fermion lines pose
a comparatively simpler problem. In any case, Chern-Simons theory has
little to add to our understanding of hard thermal loops with external
fermions. I shall therefore confine myself to the case of external gluon
lines.) The remarkable fact, which I hope to convince you of,
is that $\Gamma [A]$ is essentially
given by the eikonal for a Chern-Simons gauge theory. Later, I shall
introduce an auxiliary field $G$ and write an action $\Gamma [A,G]$
which is closely related to a gauged Wess-Zumino-Novikov-Witten (WZNW)
theory. The elimination of the auxiliary field $G$ will lead us back to
$\Gamma [A]$.

The starting point of my discussion will be the following two key
properties of $\Gamma [A]$ which one can show by power-counting analyses
of hard thermal loops $^{3,4,5}$.

1) $\Gamma [A]$ is gauge invariant.

2) $\Gamma [A]$ has the form
$$
\Gamma [A]~= (N+{\textstyle {1\over 2}} N_F ) {T^2 \over 12 \pi}\left[
\int d^4x~ 2\pi ~A^a_0 A^a_0 ~+~ \int d\Omega~ W(A\cdot Q)\right]
\eqno(1)
$$
We consider an $SU(N)$ gauge theory with $N_F$ flavors of quarks.
The gauge potential $A_\mu = (-it^a)A^a_\mu. ~t^a$ are $N\times N$
traceless hermitian matrices with ${\rm Tr}(t^a t^b)= {1\over
2}\delta^{ab}$; they are a basis of the Lie algebra of $SU(N)$ in the
fundamental representation. $Q^\mu$ is a null vector, $Q_\mu Q^\mu =0$;
we will parametrize $Q^\mu$ as $(1, {\vec Q})$ with ${\vec Q}\cdot {\vec
Q}=1$. The $d\Omega$-integration in Eq. (1) is over all orientations of
the unit vector ${\vec Q}$. The first term in Eq. (1) is a mass term for
time-component $A^a_0$; it is essentially the Debye screening effect.
The significant feature of Eq. (1) is that the second term of $\Gamma [A]$
is given by a functional of $A\cdot Q$; there is integration over all the
orientations of ${\vec Q}$ but for each ${\vec Q}$, only one component of the
potential, viz. $A\cdot Q$, enters. The $d\Omega$-integration is
part of the loop-integration; we carry out the $p_0$- and $\vert {\vec p}
\vert $- integrations of the loop-momentum $p_\mu$, leaving the angular
integrations.

The gauge invariance and special structure of $\Gamma [A]$ as in Eq. (1)
have been analyzed by many people $^{3,4,5}$.
I shall take these properties of $\Gamma [A]$
as the premise of my discussion. Because of the special structure of
$\Gamma [A]$, gauge invariance suffices to determine $W(A\cdot Q)~^4$. In
$\Gamma [A]$, as given by Eq. (1), we can carryout a gauge transformation
$A_\mu \rightarrow A_\mu +D_\mu \omega,~~\omega =(-it^a )\omega^a(x)$.
The variation of the terms in Eq. (1) are
$$\eqalignno{
\delta \int 2\pi A^a_0 A^a_0 ~&= \int 4\pi \partial_0 A^a_0 \omega^a
= \int d\Omega~Q\cdot \partial_0 A^a_0 \omega^a &(2a)\cr
\delta \int W(A\cdot Q)~&= -\int \left( Q\cdot \partial {\delta W\over
\delta (A\cdot Q)}~+ [A\cdot Q, {\delta W\over \delta (A\cdot Q)}]\right)^a
\omega^a &(2b)\cr}
$$
The property of gauge invariance of $\Gamma [A]$ then becomes the following
equation for $W(A\cdot Q)$.
$$
{\partial f \over \partial u}~+~ [A\cdot Q, f] ~+~ {\textstyle {1\over 2}}
{\partial (A\cdot Q)\over \partial v}=0 \eqno(3)
$$
where $u= {1\over 2}Q' \cdot x,~ v={1\over 2}Q\cdot x,~Q'^\mu =
 (1,-{\vec Q})$ and
$$
f= {1\over 2} {\delta W\over \delta (A\cdot Q)} + {\textstyle{1\over2}}
A\cdot Q \eqno(4)
$$
We also define $A_{+}= {1\over 2} A\cdot Q$. It is also convenient to do
a Wick rotation (to the Euclidean space ${\bf R}^4$); we then have
$u \rightarrow z,~v\rightarrow {\bar z}, ~A_{+}\rightarrow A_z$. Let us also
rename $-f$ as $a_{\bar z}$. Eq. (3) then becomes
$$
\partial_{\bar z} A_z ~-~\partial_z a_{\bar z} ~+~ [a_{\bar z}, A_z]=0 \eqno(5)
$$
and
$$
a_{\bar z}\equiv -f = -{1\over 2}{\delta W\over \delta A_z}-A_z \eqno(6a)
$$
or equivalently
$$
\delta W= 4 \int {\rm Tr} (a_{\bar z} \delta A_z ) - \delta \int A^a_z A^a_z
\eqno(6b)
$$

We must solve Eq. (5) for $a_{\bar z}$ in terms of $A_z$, use this in Eq. (6b)
and solve for $W$. If for a moment, we think of $(A_z, a_{\bar z})$ as the
$z$- and ${\bar z}$-components of the potential of (some other) gauge theory,
we see that Eq. (5) is the statement that the field strength $F_{z {\bar z}}$
vanishes. This is where Chern-Simons theory can give us some insights, for
Chern-Simons gauge theory is the one for which the equations of motion
say that the field strength is zero. (The mysterious renaming of variables
was to bring out this analogy.) We shall therefore keep aside Eqs. (5,6)
for the moment and have a short digression on Chern-Simons gauge theory.

Chern-Simons theory is a gauge theory in three dimensions, i.e. two spatial
dimensions. The action for the theory can be written as $^{6,7}$
$$
{\cal S}= {k\over 4\pi} \int d^3x~ {\rm Tr}\left( a_\mu \partial_\nu
a_\alpha ~+~ {\textstyle {2\over 3}} a_\mu a_\nu a_\alpha \right)
\epsilon^{\mu\nu \alpha} \eqno(7)
$$
(Here $a_\mu = (-it^a )a^a_\mu.$) We shall use complex coordinates $z=x+iy,~
{\bar z}= x-iy$ for the spatial dimensions. The equations of motion for the
action in Eq. (7) are, as I mentioned before, $F_{\mu\nu}=0$. The gauge choice
$a_0 =0$ is best suited to our purposes. In this gauge, the equations of motion
$F_{\mu\nu}=0$ become
$$
\partial_0 a_z =0,~~~~~~~~~~\partial_0 a_{\bar z}=0 \eqno(8a)
$$
$$
\partial_{\bar z} a_z - \partial_z a_{\bar z} ~+~ [a_{\bar z}, a_z ]=0
\eqno(8b)
$$
Eqs. (8a) say that the fields are independent of time; the dynamics of the
Chern-Simons theory (without sources) is trivial. The (equal-time) constraint
(8b) thus defines the theory. There are many ways to solve Eq. (8b). We shall
first take $a_z$ as the independent variable and solve for $a_{\bar z}$.
The solution is
$$
a_{\bar z}~=\sum (-1)^{n-1} \int~{d^2z_1 \over \pi}\cdots {d^2z_n \over \pi}
{}~{{a_z(1) \cdots  a_z (n) }\over {({\bar z}- {\bar z}_1 ) ({\bar z}_1 -{\bar
z}_2) \cdots ({\bar z}_n - {\bar z})}} \eqno(9)
$$
($ a_z(1)= a_z (z_1, {\bar z}_1)$, etc.) It is easy to check that this is
indeed a solution to Eq. (8b) using the identity $\partial_z {1\over {\bar z}}
= \pi \delta^{(2)}(x)$. ( This latter identity is essentially Cauchy's integral
formula. We can also interpret this as saying that ${1\over {({\bar z}-{\bar
z}')}}$ is the Green's function for $\partial_z$, convert Eq. (8b) into an
integral equation and iteratively solve it. This will lead us back
to Eq. (9). )

We now define a quantity $I(a_z )$ by the equation $^1$
$$
\delta I ~= {i\over \pi} \int d^2x~ {\rm Tr} (a_{\bar z} \delta a_z )
\eqno(10)
$$
It is trivial to check that $I(a_z)$ is given by
$$
I(a_z)~= i \sum {(-1)^n \over n}~\int {d^2z_1 \over \pi}\cdots {d^2z_n \over
\pi} ~{{ {\rm Tr} (a_z(1) \cdots a_z(n) )}\over { {\bar z}_{12} {\bar z}_{23}
\cdots {\bar z}_{n-1 n} {\bar z}_{n1}}} \eqno(11)
$$
where ${\bar z}_{ij}= {\bar z}_i-{\bar z}_j$.

This quantity $I$, mathematically defined by Eq. (10), has a very nice
interpretation in the language of analytical dynamics. In the $a_0 =0$ gauge,
the action becomes
$$
{\cal S}= ~{i\over \pi} \int d^3x~ {\rm Tr} (a_{\bar z} \partial_0 a_z ),
\eqno(12)
$$
which shows that $a_{\bar z}$ and $a_z$ are canonically conjugate variables.
Thus $\delta I$ is the analogue of $p~dx$ of point-particle mechanics. We can
integrate $p~dx$ if we express $p$ in terms of $x$. In one dimension, a
familiar constraint is the condition of fixed energy, say ${p^2 \over
2m}+V(x)=E$ or $p= {\sqrt{2m (E-V(x))}}$. The integral of $p~dx$ is the eikonal
or Hamilton's characteristic function
for the system, familiar as the exponent of the WKB wave functions in quantum
mechanics. For us, Eq. (8b) is the constraint relating the conjugate variables
$a_z$ and $a_{\bar z}$ and $I$, as defined by Eq. (10), is thus the eikonal of
the Chern-Simons theory. (The wave function of the Chern-Simons theory will be
$\Psi \sim e^{iI}$; cf. the lectures by R.Jackiw.)

It is possible to write the potential $a_z$ as $-\partial_z M~M^{-1}$, where
$M$ is a complex $N \times N$ matrix of unit determinant. (It is not
necessarily unitary.) From Eq. (8b), we then have
$a_{\bar z}= -\partial_{\bar z}M~
M^{-1}$. The eikonal $I(a_z)$ of Eq. (11) can be written in a `summed-up'
version as $I= -i {\cal S}_{WZNW}(M)$ where
$$
{\cal S}_{WZNW} = {1\over 2\pi}\int_{{\cal M}^2} {\rm Tr} (\partial_z
M~\partial_{\bar z}M^{-1} )~-{i\over 12\pi} \int_{{\cal M}^3}d^3x~
{\rm Tr}(\partial_\mu M M^{-1} \partial_\nu M M^{-1} \partial_\alpha M
M^{-1})\epsilon^{\mu\nu\alpha} \eqno(13)
$$
The action (13) defines the Wess-Zumino-Novikov-Witten theory used extensively
in studies of two-dimensional conformal field theories $^8$. The second term
in Eq. (13) involves an extension of $M$ into a three-dimensional space
${\cal M}^3$, the boundary of which is the two-dimensional world of interest
(spacetime). Eventually, physical results are independent of how this extension
is carried out. The action (13) obeys the composition rule, as can be verified
directly, $^9$
$$
{\cal S}_{WZNW} (hg)= {\cal S}_{WZNW}(h)~+~ {\cal S}_{WZNW}(g) ~-
{1\over \pi}\int_{{\cal M}^2} {\rm Tr}(h^{-1}\partial_{\bar z} h~\partial_z g
g^{-1}) \eqno(14)
$$
The infinitesimal version of this rule gives
$$
\delta {\cal S}_{WZNW}(M)= {1\over \pi} \int {\rm Tr}\left[ (\partial_{\bar z}
M M^{-1}) D_z (\delta M M^{-1})\right] \eqno(15)
$$
with $D_z \xi = \partial_z \xi +[a_z, \xi]$. This gives the identification
$I= -i{\cal S}_{WZNW}(M)$ using Eq. (10) and the expressions for
$a_z,~a_{\bar z}$ in terms of $M$.

We can now return to the plasma problem. From comparing Eqs. (5,6) to Eqs. (8b)
and (10), we see that $A_z$ now plays the role of the $a_z$ of the Chern-Simons
theory and $W$ is essentially $I$. The potential $A_z$ (and so $a_{\bar z}$)
in Eq. (5) depend on all four coordinates. However, the coordinates $x^T$
transverse to ${\vec Q}$ do not appear in Eq. (5); they just play the role of
parameters on which $A_z$ depends. This dependence is carried over to $W$ with
an integration over all $x^T$. From Eq. (11), we thus write down $W$ and
eventually $\Gamma [A]$ as
$$
\Gamma [A] = (N+{\textstyle {1\over 2}} N_F) {T^2\over 12\pi}\left[
\int d^4x~ \{2\pi A^a_0 A^a_0 -\int d\Omega (A_z^a A_z^a)
 \} -4\pi i \int d\Omega ~d^2x^T~ I(A_z)
\right] \eqno(16a)
$$
$$
I(A_z)~= i \sum {(-1)^n\over n} \int {d^2z_1\over \pi}\cdots {d^2z_n\over \pi}
{{{\rm Tr}(A_z(z_1,{\bar z}_1,x^T) A_z(z_2, {\bar z}_2, x^T)\cdots A_z(z_n,
{\bar z}_n ,x^T))}\over {{\bar z}_{12} {\bar z}_{23}\cdots {\bar z}_{n-1 n}
{\bar z}_{n1}}} \eqno(16b)
$$
(Strictly speaking $a_{\bar z}$ and $A_z$ are not complex conjugates; the
analogy holds better with a Chern-Simons theory of complex gauge group.
However, we are only using the Chern-Simons analogy to obtain Eq. (16).
The expression for $\Gamma [A]$ can also be directly checked to be a solution
to Eq. (5).)

The $n$-point functions can be directly evaluated in the kinematic
regime appropriate to hard thermal loops, at least for $n=2,3,4$. Needless to
say, the results agree with the $n$-point functions as given by Eq. (16).

Using the properties of the $d\Omega$-integration, we can write Eq. (16) as
$$
\Gamma = k \int d\Omega ~K(A_z, A_{\bar z}) \eqno(17a)
$$
$$\eqalignno{
K(A_z, A_{\bar z})&= -\left [ {1\over \pi}\int d^4x ~{\rm Tr} (A_z A_{\bar z})
+ i\int d^2x^T \left\{ I(A_z) +{\bar I}(A_{\bar z}) \right\}
\right] &(17b)\cr
&=\int d^2x^T {\cal S}_{WZNW}(M^\dagger M) &(17c)\cr}
$$
where $k=(N+{1\over 2}N_F){T^2\over 6}$.
Since the Chern-Simons action is not even under parity, its relevance to
QCD might be, at first sight, worrisome. But actually, from
Eqs. (17a) and (17b), we see
that the result is parity even. ($A_z \rightarrow A_+ , A_{\bar z}\rightarrow
A_- $ when we continue to Minkowski space.) We use Eq. (14) to obtain
Eq. (17c); the requirement that physics be independent of the extension of
$M$ into ${\cal M}^3$ usually requires that the coefficient $k$ multiplying
the action must be quantized. This is for groups with nontrivial third homotopy
group. Eq. (17c) shows that the hermitian matrix $M^\dagger M$ is what is
relevant; for hermitian matrices, the homotopy groups are trivial and there is
no requirement of quantization of $k$. We see that QCD neatly avoids what may
appear, a priori, as difficulties in using Chern-Simons or WZNW theories.
We can also write $^8$
$$
K(A_z ,A_{\bar z})= \int d^2x^T \log \det (D_z D_{\bar z}) \eqno(17d)
$$
$D_z, D_{\bar z}$ are chiral Dirac operators, which continue to
$D_+, D_-$ respectively.
For $U(1)$ gauge theory, the logarithm of the Dirac determinant, i.e.
$\log \det (D_+ D_-)$ is the mass term for the gauge boson, as we know from
 the Schwinger model. Eq. (17d) thus displays hard thermal loops or
$\Gamma [A]$, which is after all the electric mass term made gauge invariant,
as a non-Abelian generalization of Schwinger's result. (Of course, this
is on the lightcone chosen by $\vec Q$, with integration over all orientations
of $\vec Q$ eventually.)

In writing the above formulae in Minkowski space, most of the changes to be
made are obvious. The subtlety is in choosing the physically correct
$i\epsilon$-prescription in writing down the inverses of
$\partial_{\pm}~^2$.
A simple way to understand this is by writing the equations of motion
(for a $U(1)$ gauge theory) as
$$
\partial_\nu F^{\nu\mu}= iS^{-1} {\delta S \over {\delta A^{in}_\mu}}
\eqno(18)
$$
where $S$ is the scattering operator as a function of the in-fields.
This is a standard formula that goes back to the LSZ-fromulation of
field theory. If we expand out the current on the right hand side of Eq. (18),
we see that it involves multiple retarded commutators. Thus the
$i\epsilon$-prescription
we choose must give a current which has the same retardation
properties as the multiple retarded commutators.
With this understanding, we can write
the equations for the evolution of field configurations with soft momenta
as
$$
D_\nu F^{\nu\mu,a}= J^{\mu,a} \eqno(19a)
$$
$$
J^{\mu,a}= \sum_1^\infty \int {d^4k_1\over (2\pi )^4}\cdots {d^4k_n\over
(2\pi)^4} e^{-i(\sum k)\cdot x} J^{\mu,a}_n (k) \eqno(19b)
$$
$$\eqalignno{
J^{\mu,a}_n (k)= {k\over \pi} &\int d\Omega \Biggl[ {\rm Tr}
\left( ({-it^a Q^\mu
\over 2}) A_- (k_1)+ A_+ (k_1) ({-it^a Q'^\mu \over 2})\right)\delta_{n,1}\cr
&+\left\{ -(2i)^{n-1}{\rm Tr} \left( ({-it^a Q^\mu\over 2})A_+(k_1)
\cdots A_+(k_n)
\right) F(k_1, \cdots k_n)+(Q\leftrightarrow Q')\right\} \Biggr] &(19c)\cr}$$
where
$$\eqalignno{
F(k_1,\cdots k_n)&= \sum_{i=0}^n {{-q_i}\over {({\bar q}_0- {\bar q}_i)
({\bar q}_1 -{\bar q}_i)\cdots ({\bar q}_{i-1}-{\bar q}_i) ({\bar q}_{i+1}
-{\bar q}_i)\cdots ({\bar q}_n -{\bar q}_i)}}&(20a)\cr
{\bar q}_i &= \sum_{j=1}^i (k_j \cdot Q+i\epsilon_j),
{}~~~~~~~~~~~q_i = \sum_{j=1}^i k_j\cdot Q' &(20b)\cr}
$$
We may reexpress the current as
$$
J^{\nu a}= -{k\over 2\pi}\int d\Omega ~{\rm Tr} \left[ (-it^a) \left\{
(a_+ - A_+) Q'^\nu
+(Q' \leftrightarrow Q)\right\} \right] \eqno(21)
$$
where $a_+$ is defined by
$$
\partial_- a_+ - \partial_+ A_- +[ A_- ,a_+ ]=0\eqno(22)
$$
The kinetic theory calculation recently carried out by Blaizot and Iancu
also gives these equations $^{10}$.

There are two related but different contexts in which we need
the expression for $\Gamma [A]$. The first is in
setting up thermal perturbation theory. The version of $\Gamma [A]$ as
given in Eqs. (17a,b) is probably best suited for this purpose. We can
also use $\Gamma [A]$ added to the usual Yang-Mills action as an
effective action for the soft modes. (This is the spirit of Eqs. (19).)
The nonlocality of $\Gamma [A]$ makes it somewhat difficult to handle in
this context. It is useful to rewrite $\Gamma [A]$ using an auxiliary
field which makes the equations of motion local $^{11}$.
The auxiliary field we
use will be an $SU(N)$-matrix field $G(x,{\vec Q})$ which is a function
of $x$ and ${\vec Q}$, i.e. defined on ${\cal M}^4 \times S^2$,
${\cal M}^4$ being
Minkowski space. Further $G(x,{\vec Q})$ must satisfy the condition
$G^{\dagger}(x, {\vec Q})= G(x, -{\vec Q})$. The action is given by
$$\eqalignno{
{\cal S}= \int -{\textstyle{1\over 4}}F^2~+~& k\int d\Omega~\Biggl[
d^2x^T~ {\cal S}_{WZNW}(G) ~+~{1\over \pi}\int d^4x~ {\rm Tr}(
G^{-1}\partial_- G~A_+ \cr
&~-~ A_- \partial_+G~G^{-1}+A_+ G^{-1}A_- G -
A_+ A_-) \Biggr] &(23)\cr}
$$
where ${\cal S}_{WZNW}(G)$ is the WZNW action, Eq. (13), for $G$.
The quantity in the square brackets in Eq. (23) is the gauged WZNW
action $^{12}$. It is invariant under gauge transformations with $G$
transforming as $G\rightarrow G'= h(x) G~ h^{-1}(x),~h(x)\in SU(N)$.
The equations of motion are
$$
\partial_+ A_- - \partial_- a_+ ~+~ [a_+ ,A_- ]=0 \eqno(24a)
$$
$$
a_+ \equiv G A_+ G^{-1} - \partial_+ G~G^{-1}\eqno(24b)
$$
$$
(D_\mu F^{\mu\nu})^a -J^{\nu a}=0\eqno(25a)
$$
$$\eqalignno{
J^{\nu a}&= -{k\over 2\pi } \int d\Omega~ {\rm Tr}[ \{ (-it^a ) (a_+ -A_+)
Q'^\nu \} +(Q'\leftrightarrow Q)] &(25b)\cr
&= -{k\over 2\pi } \int d\Omega~ {\rm Tr}[  (-it^a) \{ G^{-1}
D_- G~Q^\nu ~-~ D_+G ~G^{-1} Q'^\nu \}] &(25c)\cr}
$$
Clearly these equations are equivalent to Eqs. (19a,21,22); the only
difference is that the equation defining $a_+$ in Eq. (22) is now
obtained as the equation of motion (24a) for $G$. Notice that the
current in Eq. (25c) looks like the current of a matter field; thus
except for the fact that $G$ depends on $\vec Q$, QCD, with hard thermal
loops added, is no stranger than Yang-Mills theory coupled to a matter
field. (One might wonder if there are solutions to Eq. (24a) which exist
even if $A_\mu =0$. These might give give extra degrees of freedom which
would vitiate the equivalence with Eqs. (19,20). Actually, from the
Minkowski version of Eq. (14) we see that ${\cal S}_{WZNW}(G)$ has
an additional gauge symmetry $G \rightarrow B(u)G~C(v)$. This is
the Kac-Moody symmetry of the WZNW-theory. This gauge symmetry
eliminates the possible `extra' solutions.)

One can also check easily that the Hamiltonian corresponding to
the action (23) is
$$
{\cal H}= \int d^3x \left[ {\textstyle {1\over 2}}(E^2+B^2)~+~
 {k\over 8\pi}\int d\Omega~{\rm Tr}\left\{ (D_0G D_0G^{-1})+ ({\vec Q}\cdot
{\vec D}G ~{\vec Q}\cdot {\vec D}G^{-1}) \right\} -A_0^a{\cal G}^a
\right] \eqno(26)
$$
where ${\cal G}^a$ is the Gauss law function, i.e. ${\cal G}^a=0$ for all
physical field configurations $^{11}$. We see that ${\cal H}\geq 0$ for such
configurations.

The auxiliary field and the positive Hamiltonian help in the analysis of
classical solutions. Clearly $G=1,~ E=B=0$ is the vacuum solution with
${\cal H}=0$. Certainly, wavelike solutions also exist. This is seen by
taking, as an ansatz, the fields to be in a $U(1)$ subgroup
of $SU(N)$,
whereupon the standard plasma wave solutions of electrodynamics are
reproduced. Presumably more general solutions, capturing the full
non-Abelian features of the theory also exist. These will be the genuine
non-Abelian plasmons of the quark-gluon plasma. Other physically
interesting possibilities are
static (perhaps unstable ) solutions corresponding to
local minima of ${\cal H}$ and nontopological solitons.

\vglue 0.6cm
\leftline{\twelvebf References}
\vglue 0.4cm
\medskip
\itemitem{1.} R.Efraty and V.P.Nair, {\twelveit Phys.Rev.Lett.}
{\twelvebf 68} (1992) 2891; {\twelveit Phys.Rev.} {\twelvebf D47} (1993)
5601.
\itemitem{2.} R.Jackiw and V.P.Nair, Columbia-MIT Report CU-TP 594,
CTP\# 2205 (to be published in {\twelveit Phys.Rev.} {\twelvebf D}).
\itemitem{3.} R.Pisarski, {\twelveit Physica A} {\twelvebf 158} (1989)
246; {\twelveit Phys.Rev.Lett.} {\twelvebf 63} (1989) 1129;
E.Braaten and R.Pisarski, {\twelveit Phys.Rev.} {\twelvebf D42} (1990)
2156; {\twelveit Nucl.Phys.} {\twelvebf B337} (1990) 569; {\twelveit
ibid.} {\twelvebf B339} (1990) 310; {\twelveit Phys.Rev.} {\twelvebf
D45} (1992) 1827.
\itemitem{4.} J.Frenkel and J.C.Taylor, {\twelveit Nucl.Phys.}
{\twelvebf B334} (1990) 199; J.C.Taylor and S.M.H.Wong, {\twelveit
Nucl.Phys.} {\twelvebf B346} (1990) 115.
\itemitem{5.} R.Kobes, G.Kunstatter and A.Rebhan, {\twelveit Nucl.Phys.}
{\twelvebf B355} (1991) 1.
\itemitem{6.} R.Jackiw and S.Templeton, {\twelveit Phys.Rev.} {\twelvebf
D23} (1981) 2291; J.Schonfeld,\break
 {\twelveit Nucl.Phys.} {\twelvebf B185}
(1981) 157; S.Deser, R.Jackiw and S.Templeton,\break
 {\twelveit
Phys.Rev.Lett.}
{\twelvebf 48} (1982) 975; {\twelveit Ann.Phys.} {\twelvebf 140} (1982)
372.
\itemitem{7.} E.Witten, {\twelveit Commun.Math.Phys.} {\twelvebf 121}
(1989) 351.
\itemitem{8.} E.Witten, {\twelveit Commun.Math.Phys.} {\twelvebf 92}
(1984) 455; V.G.Knizhnik and \break
A.B.Zamolodchikov,
 {\twelveit Nucl.Phys.}
{\twelvebf B247} (1984) 83; D.Gepner and E.Witten, {\twelveit Nucl.Phys.
} {\twelvebf B278} (1986) 493.
\itemitem{9.} A.Polyakov and P.Wiegmann, {\twelveit Phys.Lett.}
{\twelvebf 141B} (1984) 223; D.Gonzales and A.N.Redlich, {\twelveit
Ann.Phys.} {\twelvebf 169} (1986) 104; G.Dunne, R.Jackiw and
\break
C.A.Trugenberger, {\twelveit Ann.Phys.} {\twelvebf 194} (1989) 194.
\itemitem{10.} J.P.Blaizot and E.Iancu, {\twelveit Phys.Rev.Lett.}
{\twelvebf 70} (1993) 3376; Saclay Report SPHT-93-064.
\itemitem{11.} V.P.Nair, Columbia Report CU-TP 601 (to be published in
{\twelveit Phys.Rev.} {\twelvebf D}).
\itemitem{12.} R.I.Nepomechie, {\twelveit Phys.Rev.} {\twelvebf D33}
(1986) 3670; D.Karabali, Q-H.Park, \break
H.J.Schnitzer and Z.Yang, {\twelveit
Phys.Lett.} {\twelvebf 216B} (1989) 307; D.Karabali and H.J.Schnitzer,
{\twelveit Nucl.Phys.} {\twelvebf B329} (1990) 649; K.Gawedzki and
A.Kupianen, {\twelveit Phys.Lett.} {\twelvebf 215B} (1988) 119;
{\twelveit Nucl.Phys.} {\twelvebf B320} (1989) 649.

\bye